\documentclass[12pt, a4paper]{article}
\usepackage[left=2.25cm,right=2.25cm,top=1.65cm,bottom=1.65cm]{geometry} 

\usepackage{hanging}
\usepackage{graphicx}
\usepackage{authblk}

\usepackage{fontspec}
\usepackage{lineno}
\usepackage{url}

\usepackage{tikz}
\usepackage{everypage}

\usepackage{titling}
\usepackage{fontspec}
\usepackage{comment}
\newfontfamily\mytitlefont[Path=./]{ICCTITLE.ttf}
\pretitle{\begin{flushleft}\LARGE\mytitlefont}
\posttitle{\par\end{flushleft}\vskip 24pt}
\preauthor{\begin{flushleft}\hskip 70pt \normalsize}
\postauthor{\par\end{flushleft}\vskip 0em}
\usepackage{titlesec}
\titleformat{\section}{\normalsize\bfseries\uppercase}{\thesection}{1em}{}
\titlespacing{\section}{0pt}{18pt}{6pt}
\titleformat{\subsection}{\normalsize\bfseries}{\thesubsection}{1em}{}
\titlespacing{\subsection}{0pt}{6pt}{6pt}
\newcommand{\periodafter}[1]{#1.}
\titleformat{\subsubsection}[runin]
{\normalsize\bfseries}
{\thesubsubsection}{1em}
{\hspace{2em}\periodafter}
\titlespacing{\subsubsection}{0pt}{3pt}{3pt}

\usepackage[sorting=none, backend=bibtex, style=numeric, citestyle=numeric]{biblatex}
\DeclareFieldFormat{labelnumberwidth}{#1\adddot\midsentence}
\DeclareNameAlias{default}{last-first}

\DeclareBibliographyDriver{incollection}{%
  \printnames{author}%
  \setunit{\adddot\space}%
 \setunit{\addcomma\space}%
 \printtext[quotes]{ \printfield{title}}%
  \setunit{\addcomma\space}%
  \printfield{booktitle}%
    \setunit{\addcomma\space}%
  \printlist{publisher}%
  \setunit{\space}%
  \printtext[parens]{\printfield{year}}%
  \setunit{\addcomma\space}%
  \printfield{pages}%
  \finentry
}

\DeclareFieldFormat[report]{title}{#1}
\DeclareBibliographyDriver{report}{%
 \textit{\printfield{title}}%
\setunit{\labelnamepunct}\newblock
   \setunit{\addcomma\space}%
  \printnames{author}%
  \setunit{\addcomma\space}%
\printtext{Tech. Rept.}  
\setunit{\space}%
\printfield{number}%
  \setunit{\space}%
\printtext[parens]{\printfield{year}}%
  \finentry
}

\DefineBibliographyStrings{english}{
  volume = {Vol\adddot}
}
\DeclareBibliographyDriver{inproceedings}{%
  \printnames{author}%
 \setunit{\adddot\space}%
 \setunit{\addcomma\space}%
 \printtext[quotes]{ \printfield{title}}%
  \setunit{\addcomma\space}%
  \printfield{booktitle}%
  \setunit{\addcomma\space}%
    \printfield{volume}%
  \setunit{\adddot}%
     \iffieldundef{number}{}{\printfield{number}}%
 \setunit{\addcomma\space}%
 \printlist{organization}%
\setunit{\space}%
  \printtext[parens]{\printfield{year}}%
  \setunit{\addcomma\space}%
  \printfield{pages}%
  \finentry
}

\DeclareFieldFormat[article]{volume}{#1} 
\DeclareFieldFormat[article]{number}{#1} 
\DeclareBibliographyDriver{article}{%
  \printnames{author}%
  \setunit{\space}%
  \setunit{\addcomma\space}%
   \printtext[quotes]{ \printfield{title}}%
\setunit{\addcomma\space}%
  \printfield[journaltitle]{journaltitle}%
 \setunit{\space}%
   \printfield{volume}%
  \setunit{\adddot}%
     \iffieldundef{number}{}{\printfield{number}}%
 \setunit{\space}%
\printtext[parens]{\printfield{year}}%
\setunit{\addcomma\space}%
  \printfield{pages}%
  \finentry

}

\DeclareBibliographyDriver{book}{%
  \printnames{author}%
  \setunit{\adddot\space}%
 \setunit{\addcomma\space}%
 \printtext[it]{ \printfield{title}}%
  \setunit{\addcomma\space}%
  \printlist{publisher}%
  \setunit{\space}%
  \printtext[parens]{\printfield{year}}%
  \finentry
}

\usepackage{indentfirst}

\usepackage{subcaption}
\DeclareCaptionFormat{custom}
{%
    \textbf{#1#2}\textrm{\small #3}
}
\usepackage{caption}
\makeatletter
\def\@eqnnum{{\normalfont \normalcolor (\theequation)\hspace{5mm}}}
\makeatother

\usepackage[hidelinks]{hyperref}

\begin{document}

\title{Sub-Kelvin Cryogenics for a Super-Pressure Balloon-Borne CMB Polarimeter: Taurus}

\author{%
\makebox[\textwidth][l]{%
\hspace*{70pt}%
\begin{minipage}{\dimexpr\textwidth-70pt\relax}
\raggedright
\textbf{%
Jared L. May$^{a}$,
Alexandre E. Adler$^{b,c}$,
Jason E. Austermann$^{d}$,
Steven J. Benton$^{e}$,
Rick Bihary$^{a}$,
Shannon Duff$^{d}$,
Malcolm Durkin$^{f,d}$,
Jeffrey P. Filippini$^{g}$,
Aurelien A. Fraisse$^{e}$,
Thomas Gascard$^{h}$,
Sho M. Gibbs$^{g}$,
Suren Gourapura$^{e}$,
Jon E. Gudmundsson$^{h,i}$,
Johannes Hubmayr$^{d}$,
William C. Jones$^{e}$,
Ashesh Khatua$^{h}$,
Darby McCauley$^{j}$,
Johanna M. Nagy$^{a}$,
Ivan Padilla$^{a}$,
Ricardo R. Rodriguez$^{a}$,
John E. Ruhl$^{a}$,
M. Shaaf Sarwar$^{a}$,
Simon Tartakovsky$^{e}$,
Joseph van der List$^{a}$,
Michael R. Vissers$^{d}$,
Philippe Voyer$^{e}$%
}
\end{minipage}%
}%
}

\affil{%
\vskip -6pt
\makebox[\textwidth][l]{%
\hspace*{70pt}%
\begin{minipage}{\dimexpr\textwidth-70pt\relax}
\normalsize
$^{a}$Department of Physics, Case Western Reserve University, Cleveland, OH, USA\\
$^{b}$Department of Physics, University of California, Berkeley, CA 94720, USA\\
$^{c}$Computational Cosmology Center, Lawrence Berkeley National Laboratory, Berkeley, CA 94720, USA\\
$^{d}$National Institute of Standards and Technology, Boulder, CO, USA\\
$^{e}$Department of Physics, Princeton University, Jadwin Hall, Princeton, NJ, USA\\
$^{f}$Department of Physics, University of Colorado Boulder, Boulder, CO, USA\\
$^{g}$Department of Physics, University of Illinois Urbana-Champaign, Urbana, IL, USA\\
$^{h}$Science Institute, University of Iceland, Reykjavik, Iceland\\
$^{i}$The Oskar Klein Centre, Department of Physics, Stockholm University, AlbaNova, Stockholm, Sweden\\
$^{j}$Department of Astronomy, University of Illinois Urbana-Champaign, Urbana, IL, USA
\end{minipage}%
}%
}

\date{}
\maketitle
\thispagestyle{empty}


\section*{\vskip -30pt Abstract}
Taurus is a balloon-borne cosmic microwave background (CMB) experiment designed to operate more than 10,000 transition-edge sensor bolometers at a base temperature near 100~mK during a multi-week stratospheric balloon flight. This platform provides near-space observing conditions while imposing stringent constraints on mass, power, and system robustness, driving the need for a lightweight and highly reliable cryogenic system.

To meet these requirements, Taurus employs a multi-stage cryogenic architecture. A 660~L liquid helium tank provides a stable 4~K reservoir, with vapor-cooled shields establishing intermediate stages at approximately 40~K and 80~K. A superfluid helium tank provides a $\sim$1.5~K takeoff point for the sub-Kelvin (sub-K) cooling systems. Each of the instrument's three receivers is supported by an independent sub-K cooling chain that includes closed-cycle \textsuperscript{3}He sorption refrigerators that cool to 300~mK. These provide the thermal intercept and takeoff for a Chase Research Cryogenics miniature dilution refrigerator that cools the detectors to approximately 100~mK. Here we discuss the requirements and challenges of the Taurus sub-K cryogenic system and present results of early performance tests.

\section*{Introduction}

\begin{figure}[t!]
\centering
\includegraphics[scale = 0.800 , trim = 0cm 0cm 0cm 0cm, clip]{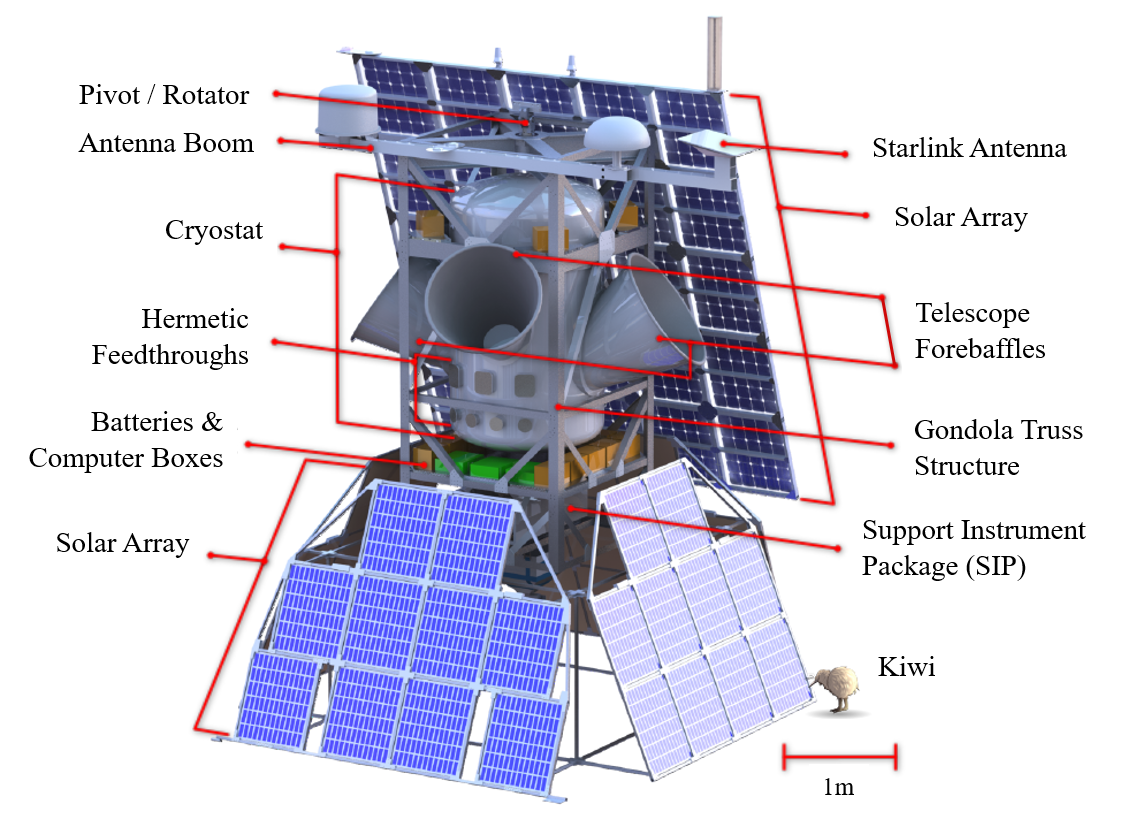}
\fontsize{11}{13}
\selectfont
\caption{A model of the Taurus instrument showing the cryostat integrated within the gondola support structure. The surrounding solar arrays, communication systems, batteries, and support instrumentation live outside of the cryostat but are critical to the payload's multi-week operation.  The receivers and sub-K cooling structures described in this paper are housed inside the cryostat. The full assembly is approximately 6~m tall.}
\fontsize{12}{14}
\selectfont
\label{fig:payload_cad}
\end{figure}

Taurus is a long-duration balloon-borne cosmic microwave background (CMB) polarimeter designed to observe from the stratosphere using NASA's super-pressure balloon (SPB) platform \cite{cathey2017performance}. The main goals are to improve measurements of a cosmological parameter known as the optical depth to reionization ($\tau$), search for cosmic anomalies, and characterize Galactic foregrounds (e.g.\ \cite{qin2020reionization, allison2015towards}). The instrument will map $\sim$70\% of the sky using more than 10,000 transition-edge sensor (TES) bolometer detectors. These detectors operate at a base temperature near 100~mK and cover four observing bands centered on 150, 220, 280, and 350~GHz. A model of the payload is shown in Figure \ref{fig:payload_cad} and is described in more detail in \cite{may2024instrument}. 

The stratospheric observing platform offers a number of advantages for CMB experiments relative to the ground, including lower atmospheric emission, access to higher microwave frequencies, and reduced large-scale atmospheric fluctuations. Taurus is targeting an April/May launch window from Wanaka, New Zealand, where the latitude and season set the day/night observing cadence. However, balloon payloads are also subject to strict constraints that the instrument must be designed to accommodate. In the context of the Taurus cryogenic system, the ballooning platform poses several additional requirements on the sub-K coolers as later discussed.

The Taurus cryogenic architecture uses a staged thermal design to reduce heat loads from the ambient environment on the colder interior stages. A 660~L liquid helium (LHe) bath provides the primary 4~K thermal reservoir. Helium boiloff from this bath is used to extract heat from vapor-cooled shields (VCSs) to establish intermediate thermal intercepts near 80~K and 40~K.  This system is described further in \cite{tartakovsky2024thermal} and is based on the architecture used by the SPIDER experiment \cite{gudmundsson2015thermal}. A 5~L superfluid helium tank provides a 1.2-1.5~K takeoff point for the sub-K cooling systems. The ambient pressure at float altitude provides passive pumping to the SFT at approximately 10~mbar. Each of the instrument's three receivers is cooled by an independent sub-K system consisting of closed-cycle \textsuperscript{3}He sorption refrigerators and a miniature dilution fridge (mini-DR). 

Here we describe the development and status of the Taurus sub-K cryogenic systems.  This proceedings is organized as follows.  First, we discuss the requirements on the sub-K cooling system imposed by the detector operations and ballooning platform.  Then we discuss the architecture of the sub-K cooling system used in each receiver.  Results are presented from performance tests of each component, and we conclude with a discussion of operational tradeoffs and plans for future work.

\section*{Sub-K Cooling Requirements}

The main role of Taurus' sub-K cooling systems is to maintain the detector arrays in the instrument's three receivers at temperatures below 100~mK during science observations.  The major requirements for this subsystem are summarized in Table \ref{tab:TableCryoReqs}.  Due to power constraints, the instrument will observe only at night (up to 16 hours, depending on latitude), so the closed-cycle refrigerators can be re-cycled during the day without loss of observing time.  The ability to operate the sub-K coolers independently in each receiver mitigates the risk of a single-point failure and also allows their operation to be optimized for differences in loading and architecture.  Due to differences in observing frequencies and detector counts between Taurus' three receivers, each will have different photon loading, wiring heat loads, and mechanical support structures. However, these differences are small enough that a uniform cooling power budget is adopted for each receiver, with a target of 2~$\mu W$ at 100~mK.  Note that since the Taurus experiment is designed to observe at fixed elevation angle, all fridges can be mounted at their optimal orientation with respect to gravity, and there is therefore no requirement on performance as a function of tilt angle.

\begin{table}[t!]
\centering
\renewcommand{\arraystretch}{1.3}
\caption{Requirements on the Taurus sub-K cooling architecture}\label{tab:TableCryoReqs}
    \begin{tabular}{llp{2.4in}}
\hline 
        Requirement & Target & Design Driver  \\
\hline\hline
Base temperature & $\leq$100~mK & Detector operation \\
Cooling power at base temperature & 2~$\mu$W & Detector operation \\
Continuous science observation duration & $\sim$ 16 hours & Detector operation \\
Total operating time  & $\geq$50 days & Estimated LHe hold time \\
Number of receivers & 3 & Science targets \\
Automation & Full & Balloon platform \\

\hline
    \end{tabular}
\end{table}

The ballooning platform poses several additional requirements on the sub-K cryogenic system.  During flight, payload communications are intentionally limited to avoid contaminating scientific data with transmitter signals.  This leads to a requirement that all subsystems be capable of autonomous operation. Additionally, the payload must survive the stresses of launch and parachute deployment at flight termination, leading to a mechanical acceleration requirement. The overall payload has additional constraints on mass, volume, and power.  These play a significant role in the design of the cryostat and influence the sub-K cooling system through that design.  Low mass and low power consumption are design drivers of the sub-K system, even though no quantitative requirements are listed in Table \ref{tab:TableCryoReqs}. 

\section*{Sub-K Cooling Architecture}

\begin{figure}[t!]
\centering
\includegraphics[scale = 0.700 , trim = 0cm 0cm 0cm 0cm, clip]{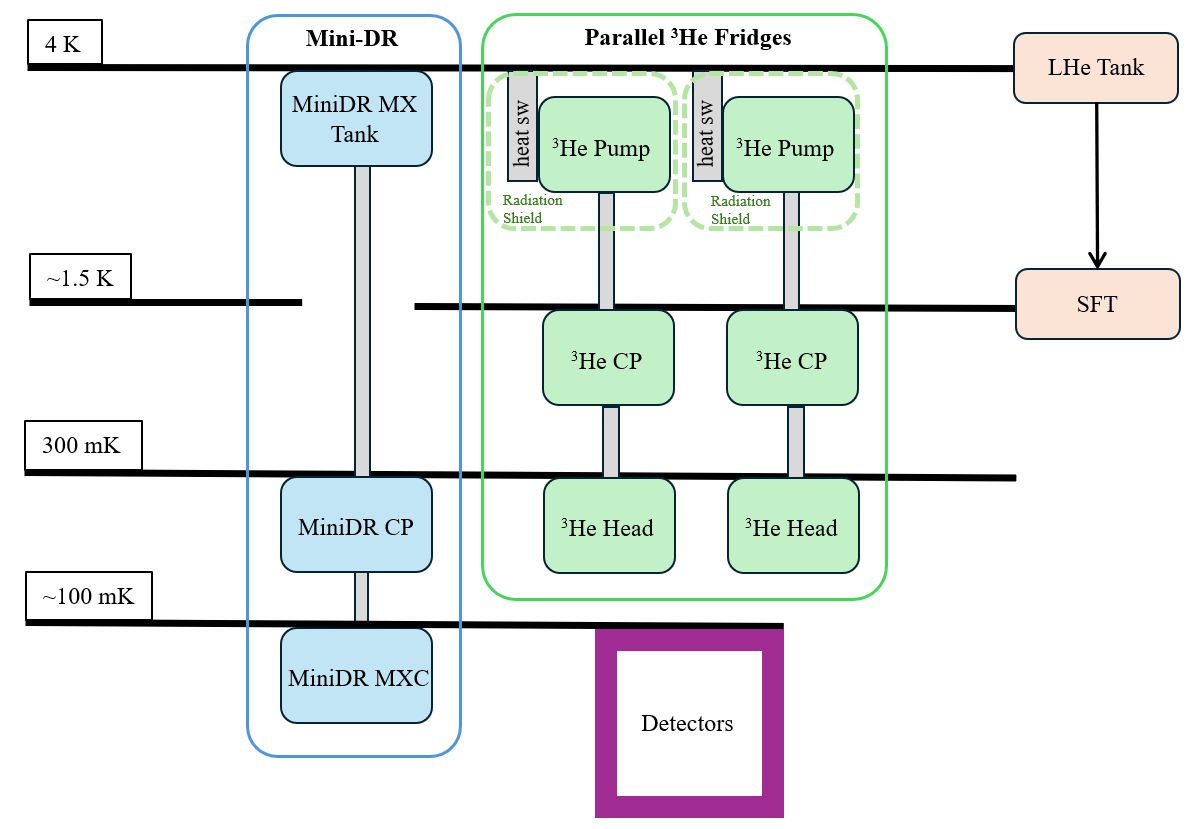}
\fontsize{11}{13}
\selectfont
\caption{A schematic diagram of the sub-K cooling system for each Taurus receiver.  The $\sim$300~mK stage is cooled by \textsuperscript{3}He sorption refrigerators.  These are thermally tied to the condensation point (CP) of the mini-DR.  The detectors are mounted to the mixing chamber (MXC) of the mini-DR, which cools them to $\sim$100~mK.}
\fontsize{12}{14}
\selectfont
\label{fig:schematic}
\end{figure}

Each Taurus receiver contains an independent sub-K cooling chain launched from the $\sim$1.5~K superfluid tank (SFT).  The sub-K cooling is provided by a combination of \textsuperscript{3}He sorption fridges that reach base temperatures near 300~mK and a mini-DR with a base temperature of $\sim$100~mK, as illustrated in Figure \ref{fig:schematic}.

The first stage of the sub-K cooling system consists of several \textsuperscript{3}He sorption fridges.  These custom fridges were fabricated by Chase Research Cryogenics~\cite{chase_research_cryogenics}, and were originally used for both flights of the SPIDER experiment \cite{gudmundsson2015thermal}. In Taurus, this $\sim$300~mK stage provides an intermediate thermal break for mechanical support structures and electrical wiring to reduce the heat load on the coldest stage. It also provides the condensation point for the mini-DR.

\begin{figure}[t!]
\centering
\includegraphics[scale = 0.500 , trim = 0cm 0cm 0cm 0cm, clip]{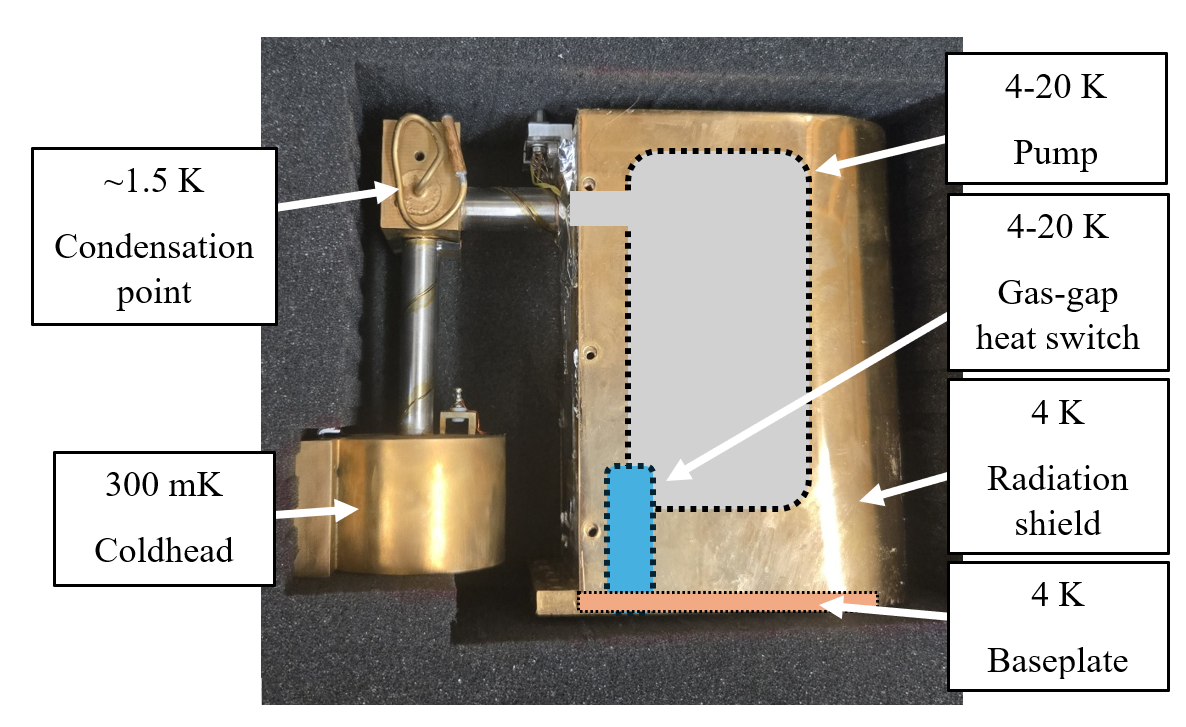}
\fontsize{11}{13}
\selectfont
\caption{Annotated view of one of the \textsuperscript{3}He sorption refrigerators used for the $\sim$300~mK stage. The photograph shows the 1.5~K condensation point and $\sim$300~mK coldhead, while the pump and gas-gap heat switch are indicated schematically behind the 4~K radiation shield. The 4~K baseplate provides the mechanical and thermal interface to the fridge.}
\fontsize{12}{14}
\selectfont
\label{fig:3He_photo}
\end{figure}

An annotated image of one of the \textsuperscript{3}He sorption fridges is shown in Figure \ref{fig:3He_photo}. Each \textsuperscript{3}He fridge features a single cooling stage, and the operation is relatively straightforward. During a fridge
cycle, heat is applied to an activated charcoal pump to release adsorbed \textsuperscript{3}He
atoms. A thermal connection to the cryostat’s $\sim$1.5~K stage provides a condensation
point for the gas, which liquefies and slowly drips into the still. When the pump heater is
turned off, closing a gas-gap heat switch provides a thermal connection to quickly cool
the pump. The pump adsorbs \textsuperscript{3}He evaporating from the still as it cools, lowering
the pressure and thus the temperature of the condensed helium to approximately 300~mK.

\begin{figure}[t!]
\centering
\includegraphics[scale = 0.650 , trim = 0cm 0cm 0cm 0cm, clip]{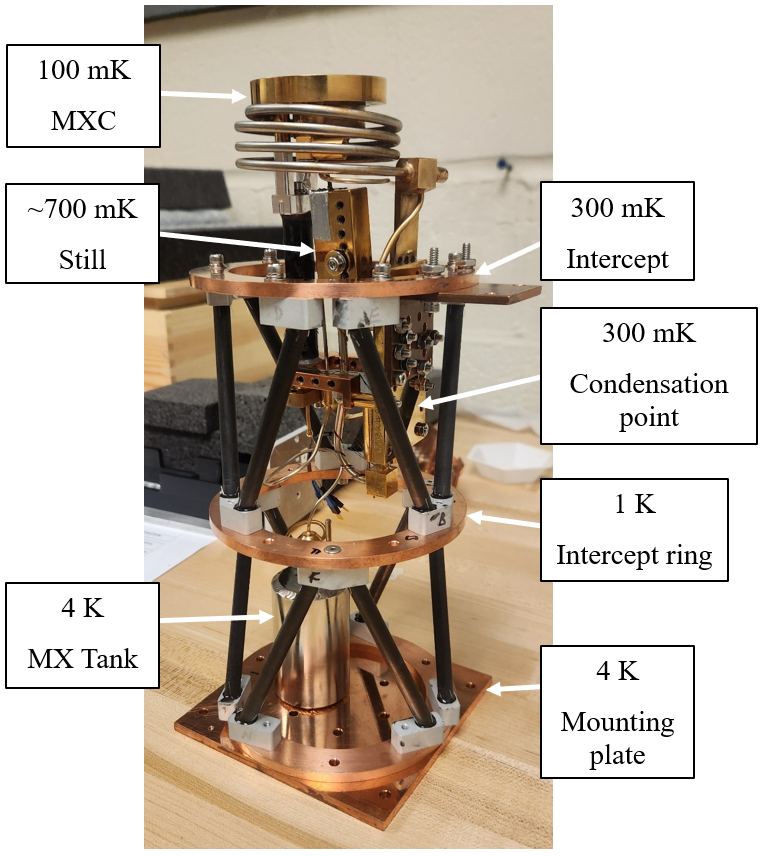}
\fontsize{11}{13}
\selectfont
\caption{Annotated photograph of the Chase Research Cryogenics mini-DR mounted in a custom support truss for lab testing. The truss provides the mechanical support for the mini-DR and establishes thermal intercepts at the 4~K, 1~K, and 300~mK stages. The mini-DR includes a 300~mK condensation point, a still operating near 700~mK, and a mixing chamber that provides the final cooling to the $\sim$100~mK base temperature.}
\fontsize{12}{14}
\selectfont
\label{fig:mini_DR}
\end{figure}

The mini-DR, also produced by Chase Research Cryogenics, provides the cooling power needed to operate the detectors at their base temperature. An annotated photograph of the mini-DR is shown in Figure \ref{fig:mini_DR}.   The design is based on the model described in \cite{teleberg2008cryogen}, but with several updates that improve overall performance. The version used by Taurus is a stand-alone mini-DR unit, without the pre-coolers sold in the standard CMD model. 

As with a typical dilution refrigerator, the mini-DR's still power is closely coupled to the base temperature of the mixing chamber. Increasing the power applied to the still increases the circulation rate in the mini-DR, which improves the cooling power at the mixing chamber and reduces the cooling time to the base temperature. The tradeoff is that the applied power at the still is transferred into the latent heat of the \textsuperscript{3}He in the mini-DR circulation and is felt directly by the \textsuperscript{3}He fridges when they condense the \textsuperscript{3}He in the mini-DR. Applying more still power thus improves the 100~mK performance but ultimately shortens the hold time of the sub-K stage and limits the length of potential continuous observing time.

Since multiple \textsuperscript{3}He fridges are used to back the mini-DR, different operating modes can be used to optimize the performance of the sub-K chain. In one configuration, the \textsuperscript{3}He fridges can be cycled simultaneously to provide more cooling power but for a shorter continuous observing time.  Alternatively, they can be cycled out of phase, so that one is regenerating while the other is providing the cooling power. In this configuration, a \textsuperscript{3}He split condenser provided by Chase Research Cryogenics is mounted between the \textsuperscript{3}He fridge coldheads and the mini-DR condensation point, effectively acting like a passive heat switch.
When cycled together, the coldheads of the \textsuperscript{3}He fridges are tied directly to the mini-DR's condensation point.  This method has the advantage of avoiding potential thermal transients caused by cycling fridges during detector operation.  On the other hand, offsetting the fridge cycles potentially enables the 300~mK stage temperature to be maintained for longer, extending the useful observing time at 100~mK.    Both methods are currently being evaluated for Taurus to assess which is a better fit for the flight operating requirements.

\section*{Preliminary Test Results}

Lab testing of the Taurus sub-K cooling system has focused on validating the performance of the \textsuperscript{3}He sorption fridges and mini-DR. These tests are intended to verify that the planned architecture can meet Taurus' requirements and to explore the tradeoffs between the different operating modes.  To facilitate rapid and cost-effective testing, the sub-K fridges are being evaluated in a pulse-tube cooled lab test cryostat rather than the Taurus flight cryostat.  Since this system does not have a superfluid helium tank, the 1.5~K cooling power is provided by a closed-cycle  \textsuperscript{4}He sorption fridge.

\begin{figure}[b!]
\centering
\includegraphics[scale = 0.750 , trim = 0cm 0cm 0cm 0cm, clip]{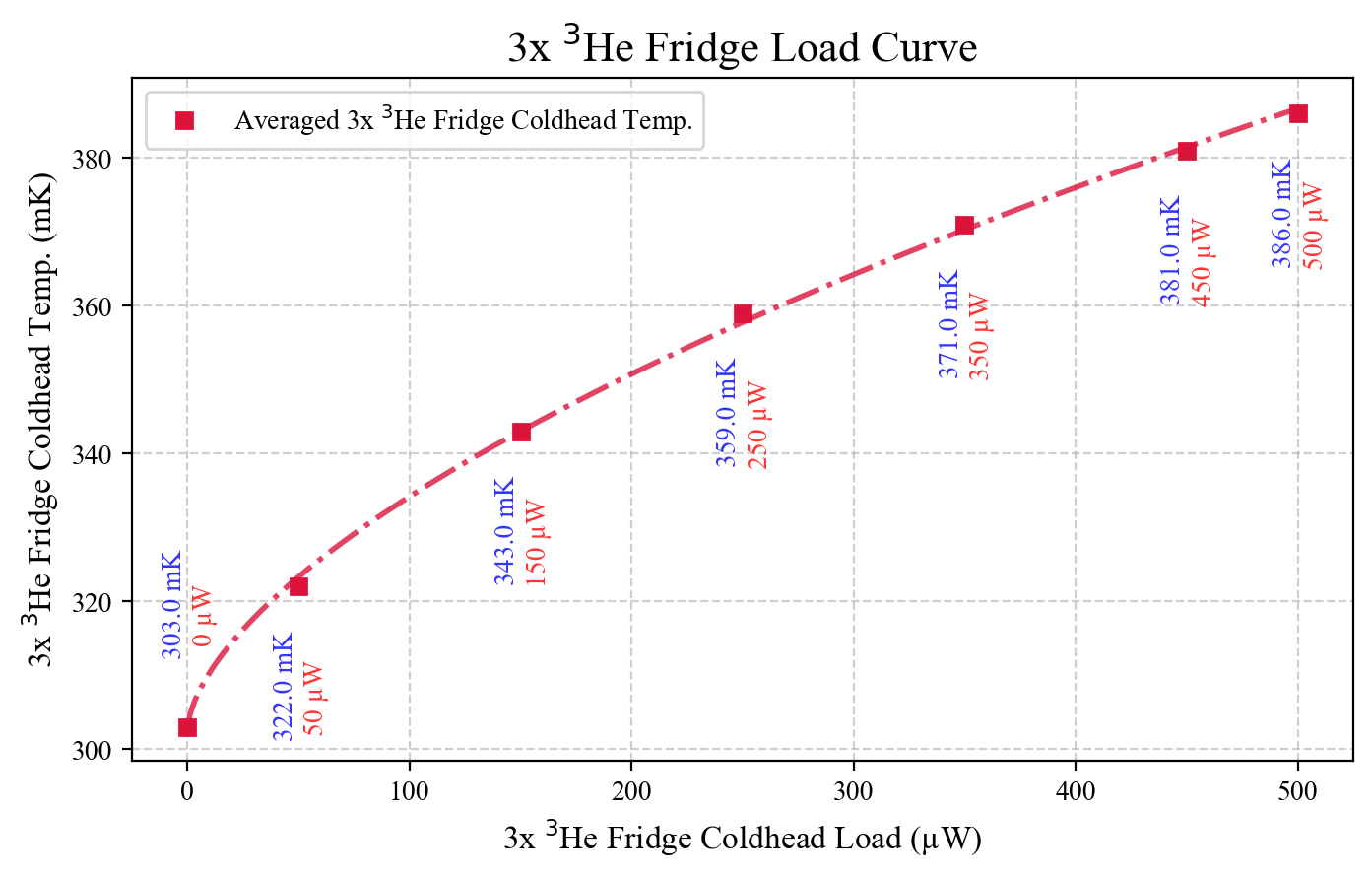}
\fontsize{11}{13}
\selectfont
\caption{Measured load curve for three \textsuperscript{3}He sorption refrigerators operated synchronously. The plotted temperature is the average coldhead temperature of the combined 300~mK stage under the applied load, providing a direct measure of the cooling performance available for the mini-DR condensation point.}
\fontsize{12}{14}
\selectfont
\label{fig:3He_loadcurve}
\end{figure}

 Tests of the \textsuperscript{3}He sorption fridges have been conducted with and without the mini-DR connected. Measurements without the mini-DR provide load curves with fewer additional thermal loads to give a better indication of \textsuperscript{3}He fridge performance.  The load curves shown in Figure \ref{fig:3He_loadcurve} show how the coldhead temperature increases with the applied load. In the Taurus receiver architecture, this load would come from a combination of the mini-DR still and the additional power from the electronic wiring and mechanical supports. The combined hold time of the fridges under a given load would also be a good indication of the hold time expected during flight, except that the lab testbed is dominated by the hold time of the $\sim$1~K sorption fridge.  Those measurements are therefore deferred to the Taurus flight cryostat, where the fridges can be tested with more realistic condensation temperatures and thermal loads.

\begin{figure}[t!]
\centering
\includegraphics[scale = 0.740 , trim = 0cm 0cm 0cm 0cm, clip]{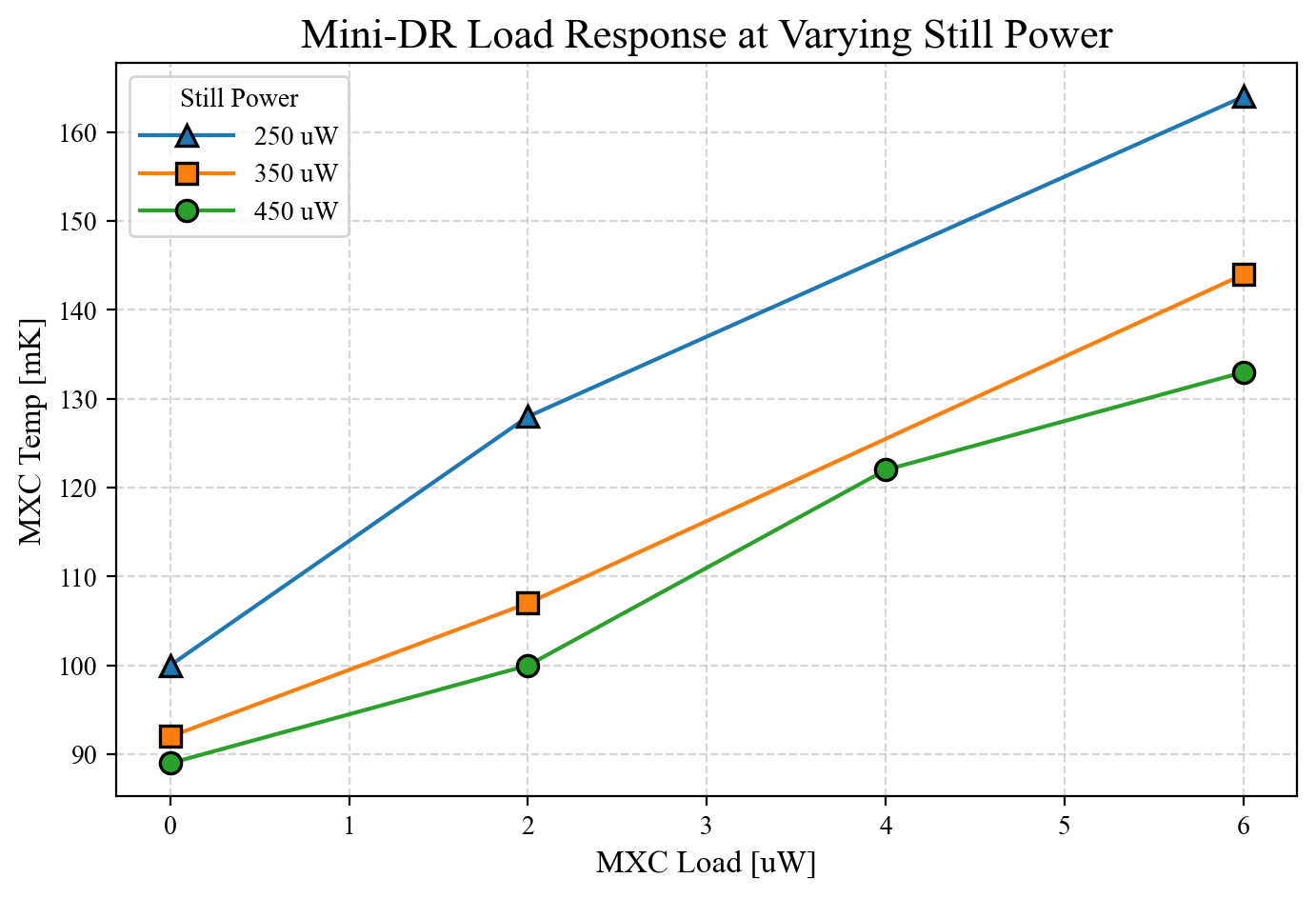}
\fontsize{11}{13}
\selectfont
\caption{Measured mini-DR load curves for several applied still powers. Increasing the still power lowers the base temperature of the mixing chamber (MXC), demonstrating improved 100~mK cooling power at the cost of an increased load on the $\sim$300~mK stage.}
\fontsize{12}{14}
\selectfont
\label{fig:miniDR_lc}
\end{figure}

Load curves have also been measured for the mini-DR in the lab testbed.  Results from tests of the mini-DR with three \textsuperscript{3}He fridges in parallel are shown in Figure \ref{fig:miniDR_lc}. This mini-DR was shown to provide 2~$\mu W$ of cooling power at 100~mK with 450~$\mu W$ applied to the still. As expected, the load curves also show a strong dependence on the applied still power. Increasing the still power increases the available cooling power at the mixing chamber (MXC) and ultimately lowers the base temperature. However, this is at the cost of an increased heat load and reduced hold time at the 300~mK fridges.  When the fridges are cycled simultaneously, this directly reduces the available continuous observing time. In this mode, the mini-DR has demonstrated base-temperature variability of $\sim$1.5~mK over 2.5 hours. The alternative out-of-phase cycling of the \textsuperscript{3}He fridges could increase the hold time and may therefore offer a promising option for achieving optimal mini-DR cooling power.  Preparations are currently underway to test this operating mode, and the measured mini-DR performance will be compared to the results with simultaneous \textsuperscript{3}He fridge operation.

\section*{Conclusions}

Taurus requires stable sub-K cooling for more than 10,000 TES bolometers during a long-duration super-pressure balloon flight. The receiver-level sub-K cooling architecture uses closed-cycle \textsuperscript{3}He sorption refrigerators to provide a $\sim$300~mK stage and miniature dilution refrigerators to cool the detectors to approximately 100~mK. This architecture allows each of the three receivers to use an independent sub-K cooling chain while sharing the common 4~K and 1.5~K cryogenic infrastructure of the payload.

Here we have described the requirements on the Taurus sub-K cooling system, the receiver-level architecture, and preliminary performance tests. Measurements of the \textsuperscript{3}He sorption fridges characterize the available $\sim$300~mK cooling performance for the mini-DR condensation point, while mini-DR load curves demonstrate Taurus-relevant cooling performance, including 2~$\mu$W of cooling power at 100~mK with 450~$\mu$W of applied still power. These results show that the component-level cooling performance is consistent with the needs of the Taurus receiver architecture.

The primary remaining question is the operating mode that provides the best 100~mK observing duty cycle. Increasing the mini-DR still power improves the cooling power at the mixing chamber, but also increases the load on the 300~mK \textsuperscript{3}He stage. Future tests will compare simultaneous and out-of-phase operation of the \textsuperscript{3}He fridges, including the use of the split condenser, to determine whether parallel single-shot operation or out-of-phase continuous operation is better suited to the Taurus flight configuration. Final validation will be performed in the Taurus flight cryostat with realistic receiver thermal loads.

\section*{Acknowledgements}
The US Taurus collaborators are supported by NASA awards 80NSSC21K1957 and \\80NSSC25K0372.  Work at the University of Iceland is supported by the Icelandic Research Fund (Grant number: 2410656-051).  The CWRU group is also supported by funding from NASA award \\80NSSC25K0609 and award CS-CSA-2024-005 from the Research Corporation for Science Advancement.

\printbibliography[title={REFERENCES}]

@inproceedings{cathey2017performance,
  title={Performance highlights of NASA super pressure balloon mid-latitude flights},
  author={Cathey, Henry M and Fairbrother, Debora A and Said, Magdi A},
  booktitle={AIAA Balloon Systems Conference},
  pages={3091},
  year={2017}
}

@inproceedings{tartakovsky2024thermal,
  title={Thermal architecture for a cryogenic super-pressure balloon payload: design and development of the Taurus flight cryostat},
  author={Tartakovsky, Simon and Adler, Alexandre E and Austermann, Jason E and Benton, Steven J and Bihary, Rick and Durkin, Malcolm and Duff, Shannon M and Filippini, Jeffrey P and Fraisse, Aurelien A and Gascard, Thomas JLJ and others},
  booktitle={Ground-based and Airborne Telescopes X},
  volume={13094},
  pages={1716--1724},
  year={2024},
  organization={SPIE}
}

@inproceedings{may2024instrument,
  title={Instrument overview of Taurus: a balloon-borne CMB and dust polarization experiment},
  author={May, Jared L and Adler, Alexandre E and Austermann, Jason E and Benton, Steven J and Bihary, Rick and Durkin, Malcolm and Duff, Shannon M and Filippini, Jeffrey P and Fraisse, Aurelien A and Gascard, Thomas JLJ and others},
  booktitle={Ground-based and Airborne Telescopes X},
  volume={13094},
  pages={1319--1333},
  year={2024},
  organization={SPIE}
}

@misc{chase_research_cryogenics,
  author       = {{Chase Research Cryogenics}},
  title        = {{Chase Research Cryogenics}},
  howpublished = {\url{https://www.chasecryogenics.com/}},
  note         = {Accessed: 2026-06-02}
}

@article{gudmundsson2015thermal,
  title={The thermal design, characterization, and performance of the SPIDER long-duration balloon cryostat},
  author={Gudmundsson, JE and Ade, PAR and Amiri, M and Benton, SJ and Bock, JJ and Bond, JR and Bryan, SA and Chiang, HC and Contaldi, CR and Crill, BP and others},
  journal={Cryogenics},
  volume={72},
  pages={65--76},
  year={2015},
  publisher={Elsevier}
}

@article{qin2020reionization,
  title={Reionization inference from the CMB optical depth and E-mode polarization power spectra},
  author={Qin, Yuxiang and Poulin, Vivian and Mesinger, Andrei and Greig, Bradley and Murray, Steven and Park, Jaehong},
  journal={Monthly Notices of the Royal Astronomical Society},
  volume={499},
  number={1},
  pages={550--558},
  year={2020},
  publisher={Oxford University Press}
}

@article{allison2015towards,
  title={Towards a cosmological neutrino mass detection},
  author={Allison, Rupert and Caucal, Paul and Calabrese, Erminia and Dunkley, Joanna and Louis, Thibaut},
  journal={Physical Review D},
  volume={92},
  number={12},
  pages={123535},
  year={2015},
  publisher={APS}
}

@article{teleberg2008cryogen,
  title={A cryogen-free miniature dilution refrigerator for low-temperature detector applications},
  author={Teleberg, Gustav and Chase, ST and Piccirillo, L.},
  journal={Journal of Low Temperature Physics},
  volume={151},
  number={3},
  pages={669--674},
  year={2008},
  publisher={Springer}
}
\end{document}